\documentclass[final]{svjour3}
\usepackage[dvipdfmx]{graphicx}
\usepackage{rotating}
\usepackage{amssymb}
\usepackage{verbatim}
\usepackage{color}
\usepackage{mathptmx}
\usepackage[numbers]{natbib}
\makeatletter
\journalname{Journal of Low Temperature Physics}
\def\bossarea{408\,${\rm deg}^2$}
\def\galaxyarea{90\,${\rm deg}^2$}
\def\exclaimz{$0 < z < 3.5$}
\def\exclaimband{420{-}540\,GHz}
\def\exclaimr{R\,{=}\,512}

\def\mspec{$\mu$-Spec}
\def\aperture{74\,cm}

\def\floattemp{1.5\,K}

\newcommand{\sups}[1]{\textsuperscript{#1}}

\bibpunct{}{}{,}{s}{}{,}
\begin{document}

\newcommand{\hdblarrow}{H\makebox[0.9ex][l]{$\downdownarrows$}-}
\title{The Experiment for Cryogenic Large-aperture Intensity Mapping (EXCLAIM)}

\author{
P. A. R. Ade\sups{9}\and 
C. J. Anderson\sups{5}\and
E. M. Barrentine\sups{7}\and
N. G. Bellis\sups{7}\and
A. D. Bolatto\sups{4}\and
P. C. Breysse\sups{1}\and
B. T. Bulcha\sups{7}\and
G. Cataldo\sups{7}\and
J. A. Connors\sups{7}\and 
P. W. Cursey\sups{7}\and 
N. Ehsan\sups{7}\and 
H. C. Grant\sups{4}\and
T. M. Essinger-Hileman\sups{7}\and 
L. A. Hess\sups{7}\and 
M. O. Kimball\sups{7}\and 
A. J. Kogut\sups{7}\and 
A. D. Lamb\sups{6}\and
L. N. Lowe\sups{7}\and 
P. D. Mauskopf\sups{8}\and 
J. McMahon\sups{10}\and
M. Mirzaei\sups{7}\and 
S. H. Moseley\sups{7}\and 
J. W. Mugge-Durum\sups{4}\and
O. Noroozian\sups{7}\and 
U. Pen\sups{1}\and
A. R. Pullen\sups{3}\and
S. Rodriguez\sups{7}\and
P. J. Shirron\sups{7}\and
R. S. Somerville\sups{2}\and
T. R. Stevenson\sups{7}\and
E. R. Switzer\sups{7}\and
C. Tucker\sups{9}\and
E. Visbal\sups{2}\and
C. G. Volpert\sups{4}\and
E. J. Wollack\sups{7}\and
S. Yang \sups{3}
}
\institute{
\sups{1} Canadian Institute for Advanced Research, CIFAR Program in Gravitation and Cosmology, Toronto, M5G 1Z8, ON, Canada \\
\sups{2} Center for Computational Astrophysics, Flatiron Institute, 162 5th Avenue, New York, NY, 10010, USA \\
\sups{3} Center for Cosmology and Particle Physics, Department of Physics, New York University, 726 Broadway, New York, NY, 10003, U.S.A. \\
\sups{4} Department of Astronomy, University of Maryland, College Park, MD 20742, USA \\
\sups{5} Department of Physics and Astronomy, Johns Hopkins University, Baltimore, MD 21218, USA \\
\sups{6} Missouri University of Science and Technology, Rolla, Missouri 65409, USA \\
\sups{7} NASA Goddard Space Flight Center, Greenbelt, MD 20771 USA \\
\sups{8} School of Earth and Space Exploration, Arizona State University, Tempe, AZ, 85287, USA \\
\sups{9} School of Physics and Astronomy, Cardiff University, Cardiff CF10 3XQ, United Kingdom \\
\sups{10} University of Michigan Department of Physics, 450 Church St, Ann Arbor, MI 48104 USA \\
$^{*}$ Corresponding author: eric.r.switzer@nasa.gov
}

%\institute{Department of Physics, Name University,\\ City, STATE zip, Country\\ Tel.:\\ Fax:\\
%\email{name@email.com}}

\titlerunning{EXCLAIM}
\authorrunning{EXCLAIM Collaboration}

\maketitle

\begin{abstract}
The EXperiment for Cryogenic Large-Aperture Intensity Mapping (EXCLAIM) is a cryogenic balloon-borne instrument that will survey galaxy and star formation history over cosmological time scales. Rather than identifying individual objects, EXCLAIM will be a pathfinder to demonstrate an intensity mapping approach, which measures the cumulative redshifted line emission. EXCLAIM will operate at \exclaimband\ with a spectral resolution \exclaimr\ to measure the integrated CO and [CII] in redshift windows spanning \exclaimz. CO and [CII] line emissions are key tracers of the gas phases in the interstellar medium involved in star-formation processes. EXCLAIM will shed light on questions such as why the star formation rate declines at $z < 2$, despite continued clustering of the dark matter. The instrument will employ an array of six superconducting integrated grating-analog spectrometers (\mspec) coupled to microwave kinetic inductance detectors (MKIDs). Here we present an overview of the EXCLAIM instrument design and status.

\keywords{Galaxy evolution, intensity mapping, MKID, integrated spectrometers}

\end{abstract}

\vspace{-5mm}
\section{Instrument Overview}
\vspace{-2mm}
Observations of our universe over cosmological time\citep{2014ARA&A..52..415M} point to an increasing rate of star formation from the period of cosmological reionization up to redshifts of $z{\approx}2$. Following this peak, the star formation rate\citep{2013ARA&A..51..105C} is thought to fall by a factor of $20$, all while dark matter continues to cluster $100$-fold. To gain a better understanding of the aggregate evolution of galaxies in this cosmological context, we need new measurements of the typical abundance, excitation, and evolution of the molecular gas that forms and is influenced by stars. The EXperiment for Cryogenic Large-Aperture Intensity Mapping (EXCLAIM) is a new high-altitude balloon spectrometer mission that will address this need by mapping the sub-millimeter emission of redshifted carbon monoxide (CO) and [CII] lines in windows comprising \exclaimz.

Instead of detecting individual galaxies, EXCLAIM will pursue intensity mapping (IM) to measure the statistics of brightness fluctuations of redshifted, cumulative line emission. IM is a measurement of integrated surface brightness rather than flux, which relaxes requirements on the EXCLAIM telescope aperture size. IM is sensitive to the integral of the luminosity function in cosmologically large volumes, and to tracers of several environments in the interstellar medium (ISM).

\begin{figure}
\begin{center}
\begin{tabular}{ll}
\includegraphics[width=0.46\textwidth]{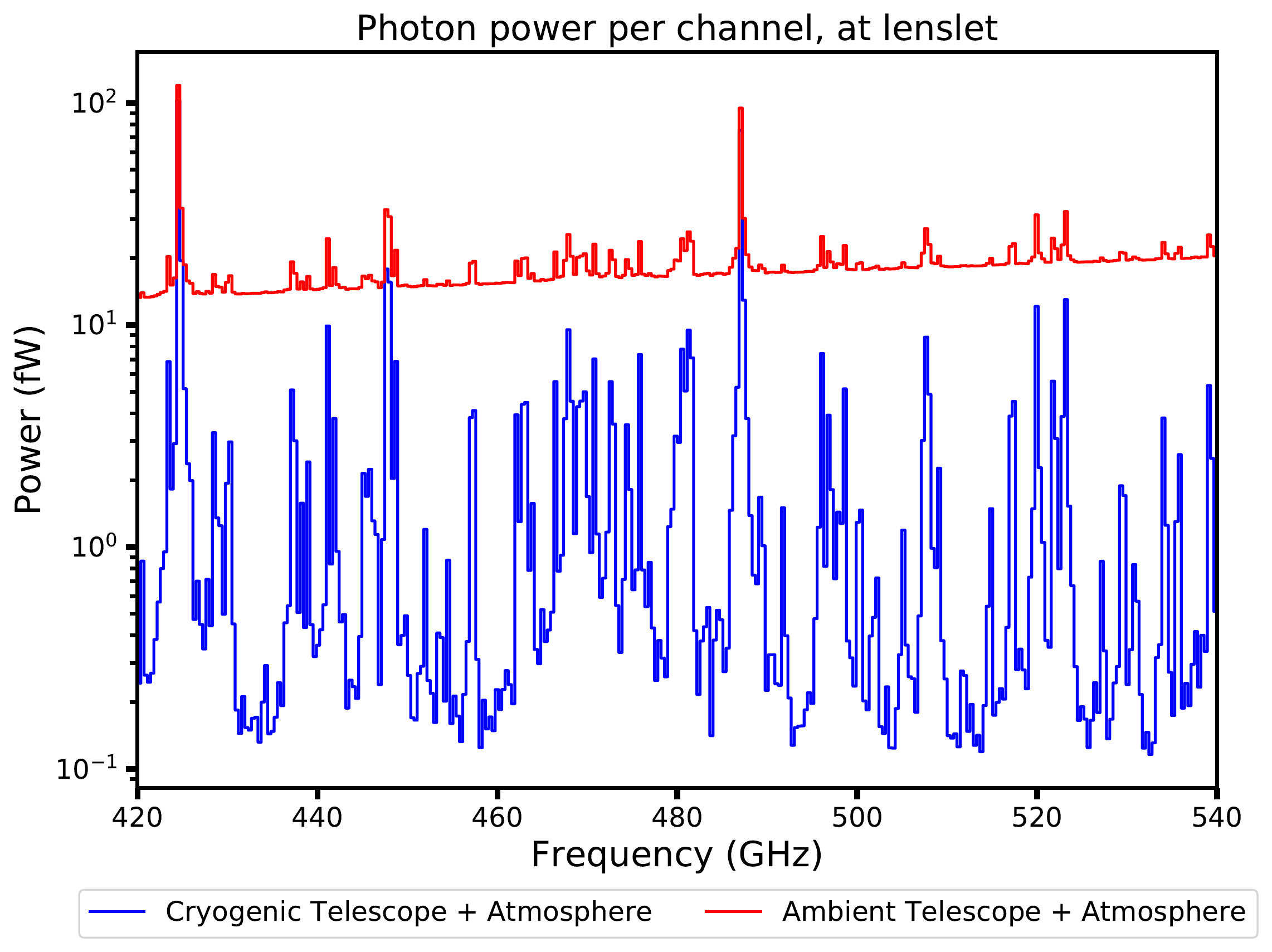} &
\includegraphics[width=0.47\textwidth]{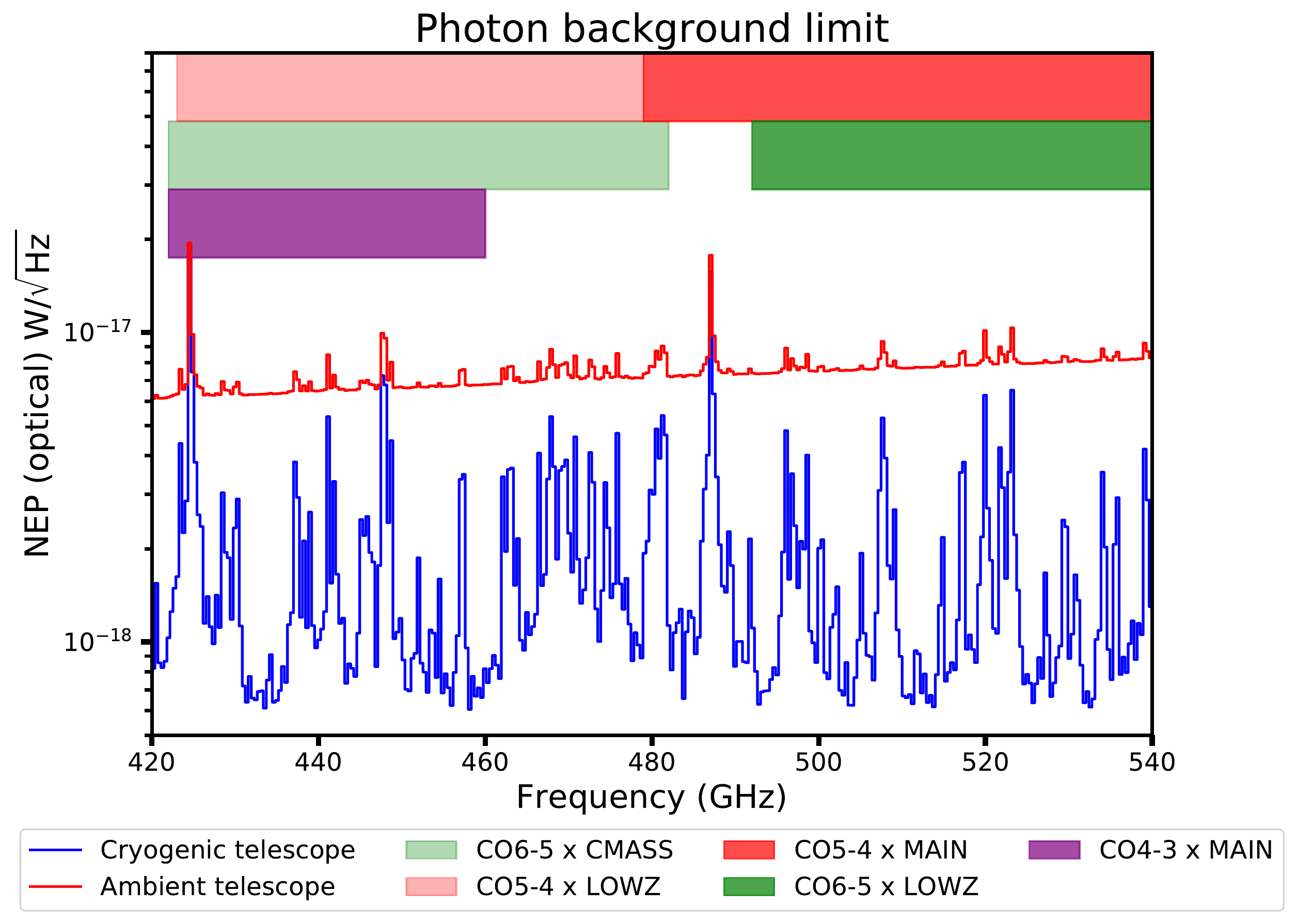}
\end{tabular}
  \end{center}
  \vspace{-10pt}
  \caption[Optics]{Photon loading (left) and background-limited optical NEP (right), both measured at the spectrometer lenslet input and assuming an overall telescope coupling efficiency of $60\%$. The background limit includes optically-excited quasiparticles and assumes $30\%$ spectrometer efficiency. The column depth and pressure broadening (${\sim}10$\,MHz/Torr) of the atmosphere drop dramatically from the ground to the altitude of balloon float. In broad photometric bands, the upper atmospheric emission is similar to one bounce from a mirror or transmission through a window at ambient temperature. The diminished pressure broadening also opens windows that are much darker than the broadband average, and which become accessible to a spectrometer with $R > 100$. Observation in these windows also benefits from cryogenic optics. In \exclaimband\ with \exclaimr, $50\%$ of the channels are $25\times$ darker than an ambient-temperature optic. \vspace{-5mm}
}
  \label{fig:photon_background}
\end{figure}

In its baseline survey, EXCLAIM will map emission across \exclaimband\ with resolving power \exclaimr\ on a \bossarea\ region of the sky overlapping with the Baryon Oscillation Spectroscopic Survey (BOSS)\citep{2016MNRAS.455.1553R} and a \galaxyarea\ region in the plane of the Milky Way. The optimal band for BOSS cross-correlation is $420{-}600$\,GHz, but we will truncate at $540$\,GHz to avoid bright ortho-water emission in the upper atmosphere at $557$\,GHz. The broad survey area provides access to linear density fluctuations, which are easier to interpret than clustering within halos. EXCLAIM's primary extragalactic science is done by cross-correlation, to facilitate unambiguous detection of redshifted emission in the presence of foregrounds.

EXCLAIM's specific goals are to 1) make a definitive detection of redshifted [CII]\citep{2019MNRAS.489L..53Y, 2019MNRAS.488.3014P} ($1900$\,GHz rest frame) in correlation with quasars at redshifts $2.5 < z < 3.5$; 2) detect two adjacent ladder lines of CO in each of the BOSS samples (MAIN, LOWZ, CMASS); and 3) constrain both CO $J=4-3$ and [CI] (492\,GHz) in the Milky Way. Current models of the diffuse, redshifted CO and [CII] intensity vary by nearly two orders of magnitude \citep{2013ApJ...768...15P, 2011ApJ...741...70L, 2008A&A...489..489R, 2010JCAP...11..016V, 2016MNRAS.461...93P}. EXCLAIM's IM measurements of two $J$ lines at the same redshift will permit new diagnostics of the ISM environment. Additionally, EXCLAIM's measurements of the integrated emission will complement interferometric observations by the Atacama Large Millimeter/submillimeter Array (ALMA) \citep{2019arXiv190309164D}. In the Milky Way, [CI] emission can track molecular gas in regions where CO is photo-dissociated. Thus, EXCLAIM will provide insight into the relation between CO and molecular gas. 

EXCLAIM will use an all-cryogenic telescope that provides the high sensitivity (see Fig.~\ref{fig:photon_background}, which describes the expected photon loading) to achieve these goals in a one-day conventional balloon flight from North America. This conventional flight provides excellent access to the BOSS regions and easy logistics and reuse. The EXCLAIM program began in April 2019 and is in a design stage. The engineering flight (with one spectrometer) is expected in 2021, and a science flight (with six spectrometers) in 2022. Here we describe the current instrument design and detection forecasts.
\vspace{-5mm}
\section{Cryogenic Telescope}
\vspace{-1mm}
\begin{figure}
\begin{center}
    \includegraphics[width=1.\textwidth]{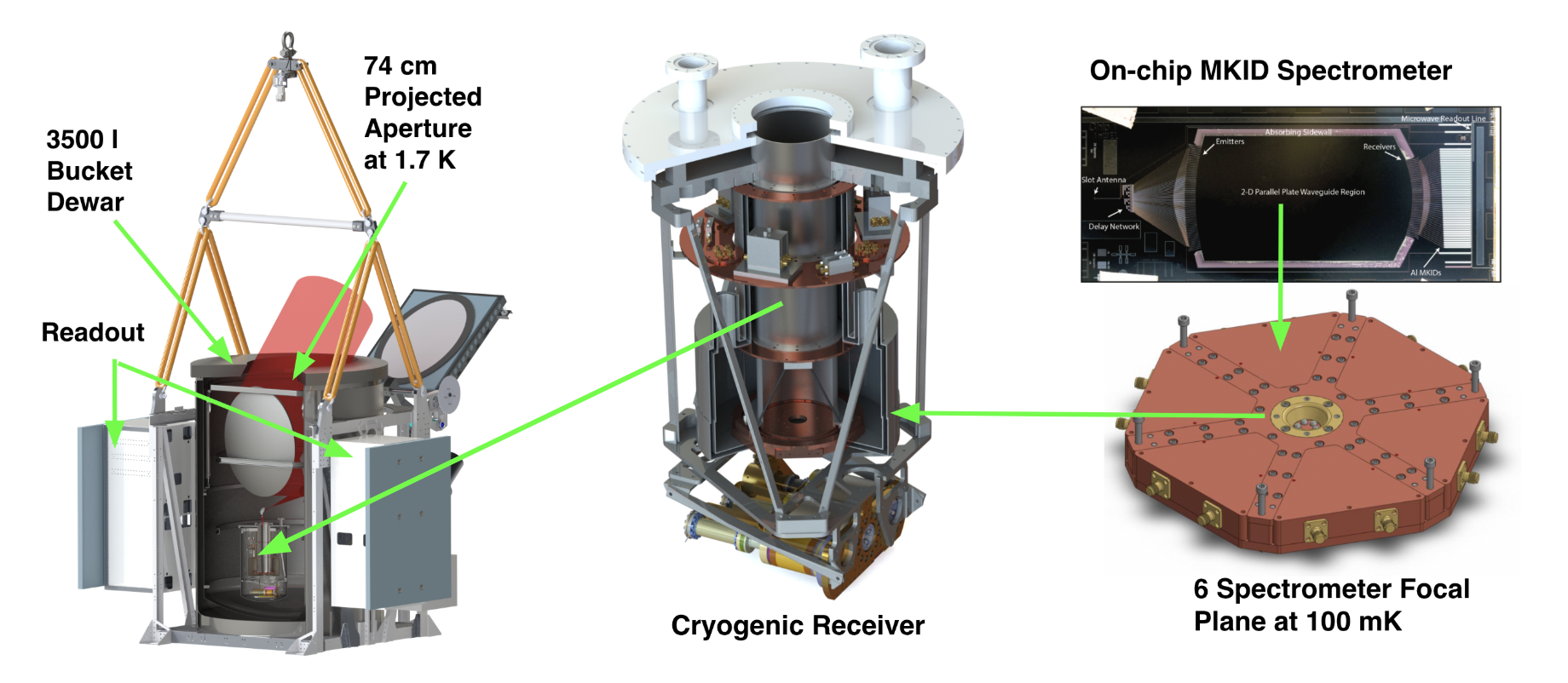}
  \end{center}
  \vspace{-15pt}
  \caption[Optics]{EXCLAIM employs a cryogenic telescope to reach high sensitivity in spectral windows of low emission in the upper atmosphere. It uses an approach with heritage from PIPER and ARCADE\,II.  A \aperture\ (projected) primary fits within the constraints of the PIPER bucket Dewar.\vspace{-5mm}}
  \label{fig:exclaim_overview}
\end{figure}

EXCLAIM's telescope and receiver will be housed in a liquid-helium Dewar, with an inner diameter of $152$\,cm, and identical in design to the Absolute Radiometer for Cosmology, Astrophysics, and Diffuse Emission (ARCADE II)\citep{2011ApJ...730..138S} and the Primordial Inflation Polarization ExploreR (PIPER) \citep{2016SPIE.9914E..1JG} instruments. An initial fill of $3500$\,l will provide $18$\,hr of \floattemp\ operation at the float altitude. A swiveling lid (Fig.~\ref{fig:exclaim_overview}) will insulate the instrument on the ground and in ascent. At float altitude, positive pressure from boil-off gas will act to keep the cryogenic optics dry and clean, eliminating the need for windows at ambient temperature. ARCADE~II and PIPER have similar open aperture areas and have demonstrated positive pressure and clean optics. A 2017 PIPER engineering flight \citep{2018SPIE10708E..06P} showed that superfluid fountain-effect pumps could maintain large telescope optical elements at \floattemp. 

EXCLAIM will observe at an elevation angle of $50^\circ$ with a $90$\,cm (\aperture\ projected) monolithic aluminum parabolic primary mirror \citep{2016SPIE.9914E..1KH, 2016SPIE.9914E..1JG}. Two powered mirrors couple the telescope to the receiver. The projected aperture size provides a resolution of $4'$~full width at half maximum (FWHM), permitting a survey that covers spatial scales from the linear regime $k \lesssim 0.1 h/{\rm Mpc}$ up to scales where shot noise dominates\citep{2019arXiv190710067B} $k \gtrsim 5 h/{\rm Mpc}$.  

\vspace{-5mm}
\section{Receiver}
\vspace{-2mm}
The receiver that houses the sub-Kelvin cooler, spectrometers, and amplifiers will remain superfluid-tight. EXCLAIM's window into the receiver will be either anti-reflective (AR)-coated\citep{quartzar} quartz (for which PIPER has demonstrated a superfluid indium seal that accommodates the coefficient of thermal expansion and does not rupture on parachute or landing shock) or AR-coated silicon. Cryogenic housekeeping harnesses and the spectrometer readout will route to $300$\,K via thin-wall stainless steel bellows.

EXCLAIM will feature six spectrometers and, unlike instruments with large focal planes, does not require high optical quality over a large field of view. This greatly simplifies the optics. The receiver optics will couple the reflective telescope to AR-coated Si hyper-hemispherical lenslets placed over the on-chip spectrometer slot antennas \citep{1993ITMTT..41.1738F, 2000stt..conf..407J}. The six spectrometers will populate a hexagon in the focal plane at $2.2 F\lambda$ spacing.
%and illuminate the primary with $9$\,dB edge taper.

A Continuous Adiabatic Demagnetization Refrigerator (CADR)\citep{2004Cryo...44..581S, 2019RScI...90i5104S} will provide $100$\,mK base temperature and $10\,\mu{\rm W}$ cooling power for the spectrometer focal plane. We will use NbTi coaxial cable\citep{2017ITAS...2800105W} and Kevlar suspensions to isolate the $100$\,mK stage from \floattemp. The high-current CADR magnet leads will be vapor-cooled and enter the receiver through a superfluid-tight feedthrough.

\vspace{-5mm}
\section{Integrated Spectrometer}
\vspace{-6mm}

\begin{figure}[htb]
\begin{center}
\includegraphics[width=0.95\textwidth]{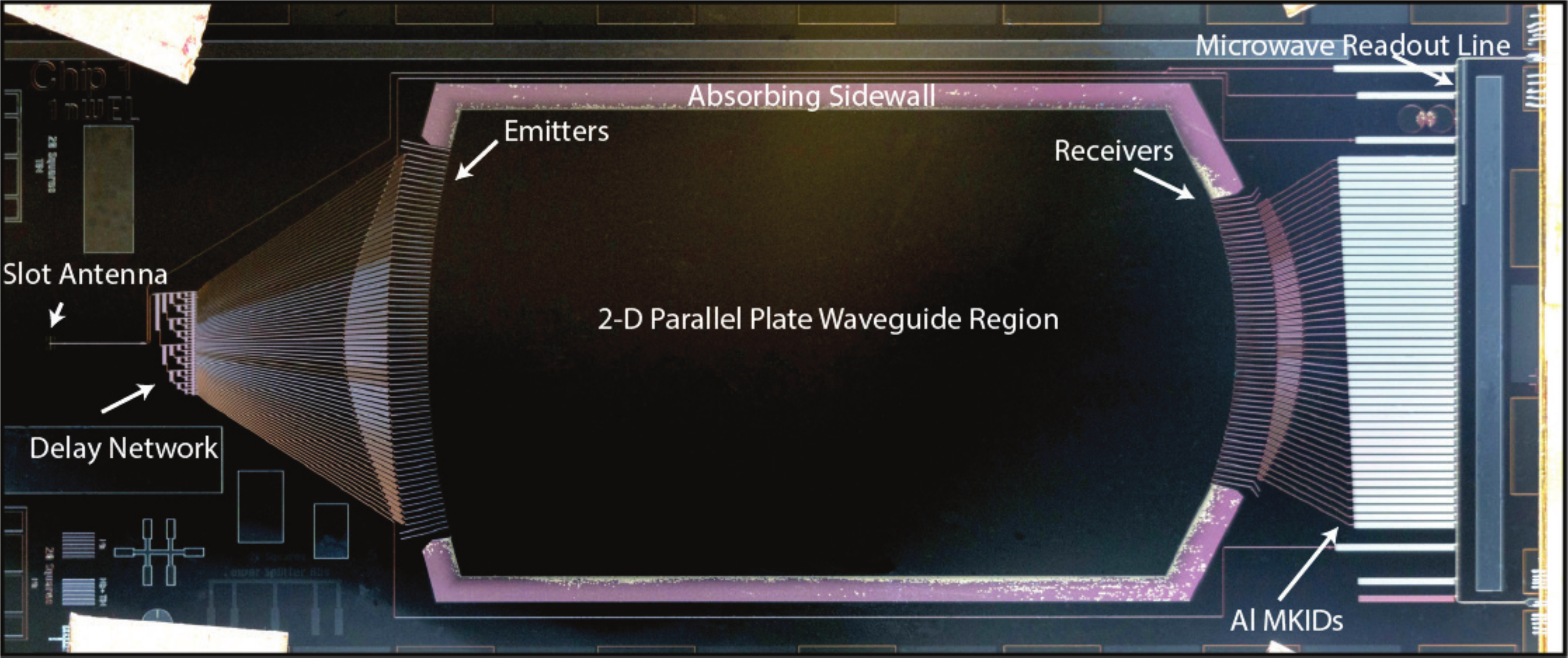}
\caption[\mspec\ device]{\label{fig:R64} Demonstrated laboratory \mspec\ prototype with R\,=\,64. This spectrometer chip is $1.8$\,cm $\times$ $4.2$\,cm in size. From the slot antenna, the light travels via superconducting Nb microstrip line to a power-splitting network. Here, in an analog of a grating spectrometer, a linear phase gradient is introduced by adjusting the length of meandered microstrip lines at each stage of the delay network. Emitter feeds then launch the phase-delayed wavefronts into a 2D parallel-plate waveguide region, where they combine constructively at an angle along the focal plane, which is dependent on wavelength, in a Rowland configuration \citep{archer1984lens,rotman1963wide}. The spectral function of the instrument follows closely a ${\rm sinc}^{2}$ function, the Fourier transform of the uniformly illuminated synthetic grating, providing excellent spectral purity. Receiver feeds are placed along this focal plane to provide Nyquist sampling of the spectral function, sending the resulting signals to individual MKID detectors. \vspace{-7mm}}
\label{fig:r64photo}
\end{center}
\end{figure}

EXCLAIM will use \mspec\ \citep{2014ApOpt..53.1094C, 2014SPIE.9143E..2CC, noroozianmu, barrentine2016design, barrentine2015overview} integrated spectrometers with resolving powers of \exclaimr\ over \exclaimband. Relative to a free-space optical grating, \mspec\ achieves a $10\times$ reduction in size by using antenna-coupled planar transmission lines to incorporate all of the elements of a diffraction-grating spectrometer, including MKID detectors, onto a silicon chip. In the \mspec\ design, the diffraction-grating is synthesized by a Nb microstrip line delay network, which launches signals into a Nb 2D parallel-plate waveguide region with emitting and receiving feeds arranged in a Rowland configuration\citep{archer1984lens,rotman1963wide}. Fig.\,\ref{fig:R64} shows an R\,=\,64 \mspec\ prototype, while the preliminary design for the EXCLAIM R\,=\,512 \mspec\ is described in \citet{2019AcAau.162..155C}. The EXCLAIM design will operate at a grating order $m=2$, and an on-chip order-sorting filter will be inserted between the slot antenna and delay network to select this order. In addition, a low-pass metal mesh filter will block radiation above $m=2$. 

Each EXCLAIM spectrometer will feature 355 half-wave Al-Nb microstrip transmission line MKIDs, with $20$\,nm-thick Al. These will be fabricated through a process that has yielded Al films in 1-layer coplanar waveguide resonator structures at thicknesses of 10-25\,nm with internal quality factors $Q_{\rm int}$ of $(1-4)\times 10^6$, and long quasi-particle lifetimes\citep{omidltd} of ${\sim}1$\,ms and Nb films\cite{HessLTD2019} in similar structures with $Q_{\rm int}$ of $(5-8)\times 10^5$. The spectrometers will be fabricated using a wafer-scale bonding technique and Nb liftoff patterning process that provides precise control of line width \citep{patel2013fabrication} on a single-crystal Si dielectric. The low-loss of single-crystal Si \citep{1986JCrGr..74..605D, 1994IJIMW..15.1181A} enables near-unity efficiency  and resolutions\citep{cataldo2018design, barrentine2016design} up to R\,=\,1500 in principle. We expect an on-chip efficiency of ${\sim}50\%$ for the EXCLAIM spectrometers. This estimate includes ${\sim}5\%$ loss through the spectrometer transmission lines due to the dielectric loss of single-crystal Si\cite{1986JCrGr..74..605D,1994IJIMW..15.1181A} and  ${\sim}52\%$ efficiency through the parallel-plate waveguide region due to limitations in our chosen receiver feed design\citep{berhanultd}. However, in our detection forecasts, we have conservatively assumed a spectrometer efficiency of $30\%$.

Except for the slot antenna, the EXCLAIM spectrometer will use a microstrip architecture, which provides immunity to cross-talk and stray light. The coupling of stray power onto the detectors will be controlled through the deposition of a thin-film absorber on the back of the spectrometer chip and thermal blocking filters \citep{4751534, 2014RScI...85c4702W} on the microwave readout lines.

\vspace{-5mm}
\section{Instrument Electronics and Flight Software}
\vspace{-2mm}
The EXCLAIM detector readout electronics baseline design will incorporate heritage from The Next Generation Balloon-borne Large Aperture Submillimeter Telescope (BLAST-TNG)\citep{2018SPIE10708E..0LL} and Far Infrared
Observatory Mounted on a Pointed Balloon (OLIMPO)\citep{2019JCAP...01..039P} missions. Intermediate frequency hardware will mix the digital to analog converter (DAC) output from a Reconfigurable Open Architecture Computing Hardware (ROACH-2) system\citep{2016JAI.....541001H} to the 2-3\,GHz MKID resonance frequency range and will mix the spectrometer output down for input to the analog to digital converter (ADC). The readout provides $512$\,MHz radio frequency (RF) bandwidth, and will read the spectrometer array at $488$\,Hz. 
% with approximately $60$\,W input power

The electronics for the CADR, cryogenic and gondola housekeeping, attitude control and measurement, power management, and telemetry interface will be identical in design to PIPER. The flight software and low-power flight computers will follow a PIPER design, demonstrated in the PIPER 2017 engineering flight.
%and in integration and test.

\vspace{-5mm}
\section{Observing Strategy}
\vspace{-2mm}

EXCLAIM will scan in azimuth at constant elevation and map a strip across a constant range of declination. Unlike the auto-power\citep{1995PhRvD..52.4307K}, a cross-power detection does not suffer a penalty toward large survey areas. Thus, we seek to maximize the survey area to access linear modes while also sampling the beam in the scan and sky drift directions. For a flight from Ft. Sumner, New Mexico, the provisional survey region encompasses Sloan Stripe 82. To constrain the atmospheric emission, we will deploy a cantilevered mass on the gondola to allow for elevation dips. Primary science observations will occur when the sun is down, but additional regions may be achievable in an anti-solar scan pattern during the day.

EXCLAIM has modest pointing requirements to conduct a large-area survey within the BOSS regions. In flight, a rotator will scan the payload about a central azimuth and a slew of several degrees in an open-loop configuration. A magnetometer will determine an instantaneous reference to the scan center azimuth. Post-flight pointing reconstruction will use data from an array of gyroscopes and accelerometers, tilt sensors, and the magnetometer to tie together star camera determinations, which will provide $\approx 3"$ pointing at $1$\- second intervals.

\vspace{-5mm}
\section{Anticipated Sensitivity}
\label{sec:forecast}
\vspace{-2mm}
\begin{figure}
\begin{center}
\begin{tabular}{ll}
\includegraphics[width=0.46\textwidth]{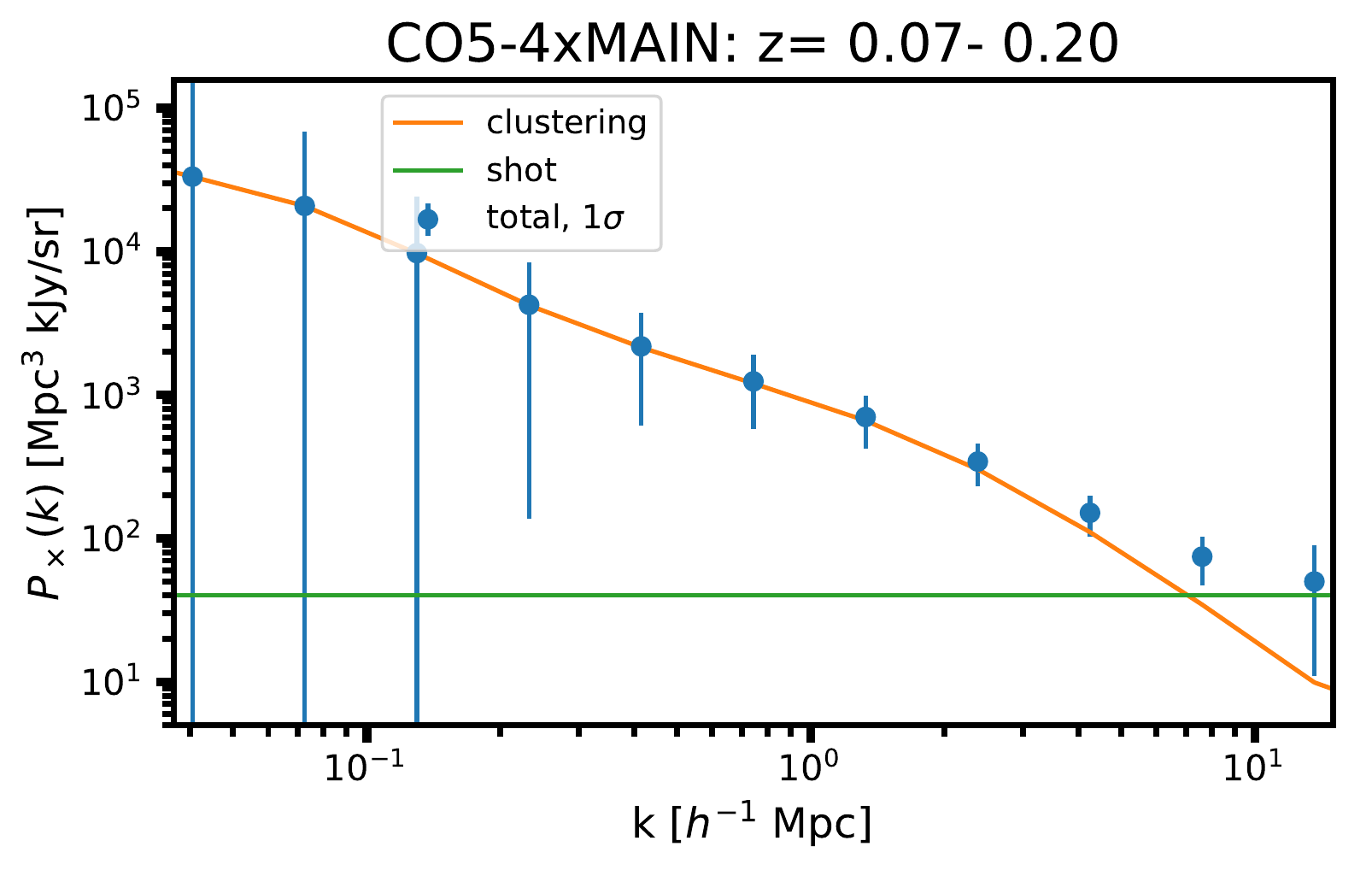} &
\includegraphics[width=0.47\textwidth]{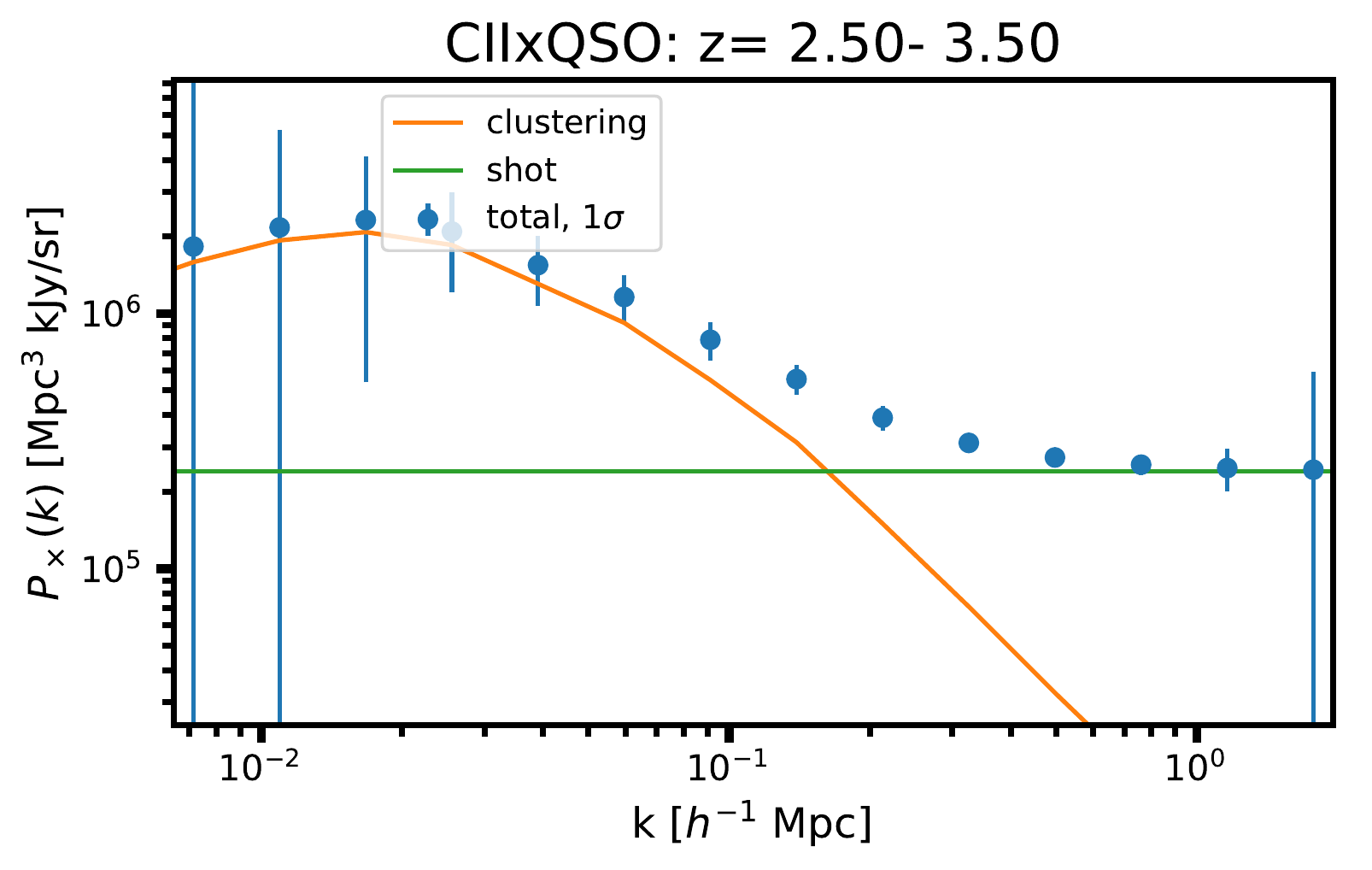}
\end{tabular}
  \end{center}
  \vspace{-10pt}
  \caption[Forecast]{Forecast for cross-correlations of CO $J=5-4$ $\times$ SDSS MAIN at $0.07<z<0.2$ (left) and [CII] $\times$ BOSS QSO at $2.5<z<3.5$ (right) assuming signal levels from \citet{2013ApJ...768...15P} and \citet{2019MNRAS.489L..53Y} respectively. The low redshift correlations trace nonlinear scales and are expected to have limited correlated shot noise. The high-redshift [CII] probes linear scales and is expected to have a higher level of correlated shot noise.
  \vspace{-5mm}
}
  \label{fig:forecast}
\end{figure}

We estimate the sensitivity for intensity mapping using both a numerically simple mode counting argument for the three-dimensional power spectrum \citep{2011ApJ...741...70L, 2014ApJ...786..111P, 2016ApJ...817..169L} and using a simulated analysis of angular cross-correlations between redshift slices \citep{2019MNRAS.485..326L}. Both estimates agree and include the effects of angular and spectral resolution and survey volume. The three-dimensional power spectrum approach requires homogeneous noise in the frequency direction, and we use an inverse-noise weighted effective noise for the volume, which is validated by simulations. Shot noise in the galaxy sample is given by published \citep{2016MNRAS.455.1553R, 2015MNRAS.453.2779E} $\bar n$. 

The specification of the detector performance must take several factors into account: 1) MKID noise contributions are loading-dependent, 2) the power absorbed by the MKID depends on the efficiency of the complete spectrometer system, and 3) MKIDs generically have $1/f$ noise  contributions from two-level systems\citep{2008ApPhL..92u2504G}. To accommodate these factors, we specify the sensitivity of the spectrometer under expected optical loading as a multiplier of the photon background-limit (including optically-excited quasiparticle fluctuations), referring to the power incident at the spectrometer lenslet, and weighted over acoustic frequencies of the science signal ($5-25$\,Hz, unless additional modulation is employed). The spectrometer NEP for these baseline forecasts is taken to be a factor of three over the background limit, though considerably poorer NEP performance will still accomplish the EXCLAIM mission threshold detection goals. As shown in Fig.~\ref{fig:forecast}, the expected $2\sigma$ sensitivity to the surface brightness-bias product for $0<z<0.2$ (SDSS MAIN) for CO J=4-3, J=5-4, $0.2<z<0.4$ for J=5-4, J=6-5 (BOSS LOWZ), $0.4<z<0.7$ for J=6-5 (CMASS), and $2.5<z<3.5$ for [CII] (QSO) are $\{ 0.15, 0.28, 0.30, 0.37, 0.45, 13 \}$\,kJy/sr, respectively.
%CO4-3xMAIN 0.151487273291 kJy / sr
%CO5-4xMAIN 0.278509387031 kJy / sr
%CO5-4xLOWZ 0.295403104812 kJy / sr
%CO6-5xLOWZ 0.36612931203 kJy / sr
%CO6-5xCMASS 0.45481422703 kJy / sr
%CIIxQSO 13.2116397966 kJy / sr

\vspace{-5mm}
\begin{acknowledgements}
We gratefully acknowledge funding provided by the NASA Astrophysics Research and Analysis (APRA) Program. 
\end{acknowledgements}

\bibliography{main}
\bibliographystyle{unsrt}

\end{document}